\documentclass[traditabstract]{aa} 

\usepackage{graphicx}
\usepackage{txfonts}
\usepackage{natbib}
\usepackage{subfig}
\bibpunct{(}{)}{;}{a}{}{,}

\newcommand{\micron}{\mathrm{\mu m}}

\begin{document}
   \title{Constraining properties of dusty environments by infrared variability}


\titlerunning{IR variability in dusty environments}

\author{S.~F.~H\"onig\inst{1} \and
M.~Kishimoto\inst{2}
}

\offprints{S.~F. H\"onig \\ \email{shoenig@physics.ucsb.edu}}

\institute{
University of California in Santa Barbara, Department of Physics, Broida Hall, Santa Barbara, CA 93106, USA \and
Max-Planck-Institut f\"ur Radioastronomie, Auf dem H\"ugel 69, 53121 Bonn, Germany
}

   \date{Received July 22, 2011; accepted September 9, 2011}

\abstract{

We present model simulations of time-variable infrared (IR) emission from dust as a consequence of variability of the incident radiation. For that we introduce a generalized treatment for temperature variations in a dusty environment, which is not limited to any specific astronomical source. The treatment has been incorporated into a simplified clumpy torus model, with the radial brightness distribution as the main parameter, to study the IR emission of type 1 active galactic nuclei (AGN). We show that any variability signal in the optical is smoothened stronger if the brightness distribution is very extended, and this smoothing strongly depends on wavelength. This also affects time lags between the optical and near-/mid-IR emission, which can be up to 10s of sublimation radii for long wavelengths and extended brightness distributions. The dependence of time lag on wavelength and distribution can be used to quantify the brightness distribution in an AGN torus, either by comparing optical light curves to near-IR and mid-IR light curves, or by directly comparing near-IR to mid-IR light curves. Moreover, our model has been applied to near-IR data of the nearby Seyfert 1 galaxy NGC~4151. We show that the simple model can reproduce the overall observed variability signal and found that about 40\% of the energy in the variability signal in the $V$-band has been converted into $K$-band variability. This low value may be explained by a ``snowball'' model of gradually-sublimating clouds at the inner edge of the torus. We also note that our modeling does not support a change of time lag/sublimation radius over the observed light curve epoch in spite of a significant change in $V$-band emission.

}
   \maketitle

%

\section{Introduction}

An optically- and geometrically-thick, circumnuclear dusty region (= ``dust torus'') is one of the key ingredients of the unification scheme of active galactic nuclei (AGN) \citep{Ant93}. It is held responsible for angle-dependent obscuration of the central accretion disk and broad-line region, explaining the apparent difference between type 1 (unobscured line-of-sight) and type 2 (obscured line-of-sight) objects. Its dust content absorbs part of the accretion disk's UV/optical emission and reemits in the infrared (IR). Since the torus resides on parsec scales, direct observations have been challenging since its angular size is much smaller than the resolution power of single telescopes in the IR.

With the advent of IR long-baseline interferometry and reverberation mapping, it is now possible to determine the sizes of the dusty region around nearby AGN \citep[e.g.][]{Jaf04,Tri07,Bec08,Kis09b,Bur09,Pot10}. Of special interest are observations of type 1 AGN where the torus is presumably seen closer to face-on than in type 2 AGN without too much complications due obscuration effects \citep{Kis09a,Kis11b}. While sizes are the primary parameter to extract from interferometry data, reverberation mapping uses the fact that any variability in the accretion disk emission causes a delayed response in the near-infrared continuum emission from the hottest, inner region of the dusty torus because of the light-travel time from the disk to the torus. Both interferometry and reverberation mapping found that the time lags and sizes are in general agreement with theoretical expectation of the sublimation radius based on thermal equilibrium of \textit{large} graphite grains or anisotropic accretion disk emission \citep[e.g.][]{Kis07}. These results were confirmed by directly measuring near-IR $K$-band ring radii using the Keck interferometer \citep{Kis09b,Kis11a}. It should be noted that variability of the accretion disk occurs on multiple time scales (from intraday variability up to many years) and amplitude ranges. The torus has been observed to react on variability in the range of few days to as long as decades \citep[e.g.][]{Gla92,Gla97,Gla04,Okn99,Gal01,Min04,Sug06}.

IR interferometry has proven as a tool to access more detailed information about the dust torus than just sizes. Based on mid-IR interferometric observations \citet{Jaf04} argued that the dust in the torus is confined to small clumps rather than smoothly distributed, as theoretically and observationally suggested in early studies \citep[e.g.][]{Kro88,Tac94}. In the following clumpy torus models have been successful in reproducing spectral \citep[e.g.][]{Nen02,Pol08,Ram11,Alo11} and spatial information \citep[e.g.][]{Hon06,Hon08,Sch08,Hon10a}. Moreover, we recently showed that the dependency of visibility on wavelength and spatial frequency can be used to constrain the radial brightness distribution of the dust \citep{Kis09b,Hon10a,Hon10b,Kis11b}. Using such an analysis, it has been found that the brightness distribution and the slope of the mid-IR spectra are strongly correlated. Both parameters are probably tracers for the radial distribution of the dust, so that combining interferometry and photometry provides a tool to characterize the properties of the dust torus in different objects.

At present IR interferometry has some tight limitations regarding brightness and spatial resolution, so that only the brightest nearby sources can be studies in detail. Reverberation mapping, on the other hand, may serve as a complimentary tool for objects that are not accessible by interferometry. In this paper we present a correspondence in reverberation mapping to the interferometric method of determining the brightness distribution in type 1 AGN. For that we use a simplified version of a clumpy torus model, which has been successfully applied to interferometric data \citep{Kis09b,Kis11b}. In Sect.~\ref{sec:var} we describe a theoretical approach to handle temperature variations in a dusty medium, generalizing the seminal work of \citet{Bar92}. Based upon this theory, we outline a simple model for type 1 AGN in Sect.~\ref{sec:varagn} and present model light curves and cross correlation function at IR wavelength that show the dependence on the brightness distribution of the torus. As a proof of concept we use our model to reproduce the $K$-band light curve of NGC~4151 in Sect.~\ref{sec:n4151}. The results are summarized in Sect.~\ref{sec:summary}.

\section{Luminosity and temperature variations in a dusty medium}\label{sec:var}

In this section we describe theoretically how variability in the luminosity of the incident radiation changes the temperature of the dust. The new dust temperatures can then be used to calculate the changed emission of the dust. Under the assumption that heating and cooling times are negligible, no iterations are required.

For dust grains in radiative equilibrium of size $a$ at distance $r$ from a radiating source with spectrum $L_\nu$, the received power per dust grain, $L_\mathrm{rec} = \int L_\nu/(4 \pi r^2) \ Q_\mathrm{abs;\nu} \ \pi a^2 \ \mathrm{d}\nu = L_\mathrm{abs} \ a^2/(4 r^2)$ equals the emitted power $L_\mathrm{em} = \int 4 \pi a^2 \ Q_\mathrm{abs;\nu} \ \pi B_\nu(T) \mathrm{d}\nu$, where $Q_\mathrm{abs;\nu}$ is the absorption efficiency of the dust and $B_\nu(T)$ denotes the Planck function of a black-body at temperature $T$. Equating $L_\mathrm{rec}$ and $L_\mathrm{em}$ leads to
\begin{equation}\label{eq:1}
L_\mathrm{abs} = 16 \pi r^2 \ Q_\mathrm{abs;P}(T) \ \sigma_\mathrm{SB} T^4\,\, ,
\end{equation}
where $\int Q_\mathrm{abs;\nu} \ \pi B_\nu(T) \ \mathrm{d}\nu$ has been replaced by the Planck mean absorption efficiency $Q_\mathrm{abs;P}(T) \cdot \sigma_\mathrm{SB} T^4$. If the emission source is varying with time, i.e. $\mathrm{d}L/\mathrm{d}t \neq 0$, the temperature of the dust grain will change. 
From Eq.~\ref{eq:1} we find
\begin{equation}\label{eq:2}
\frac{\mathrm{d}L_\mathrm{abs}}{\mathrm{d}T} = \frac{4 L_\mathrm{abs}}{T} + L_\mathrm{abs} \ \frac{Q^\prime_\mathrm{abs;P}}{Q_\mathrm{abs;P}} \,\, .
\end{equation}
Here, we define $Q^\prime_\mathrm{abs;P} \equiv \mathrm{d}Q_\mathrm{abs;P}/\mathrm{d}T$ as the $T$-derivative of the Planck mean absorption efficiency of the dust. Both the Planck mean absorption efficiency and its derivative can be calculated from the optical properties of dust and only depend on the dust temperature. 

\begin{figure}
\includegraphics[width=0.48\textwidth]{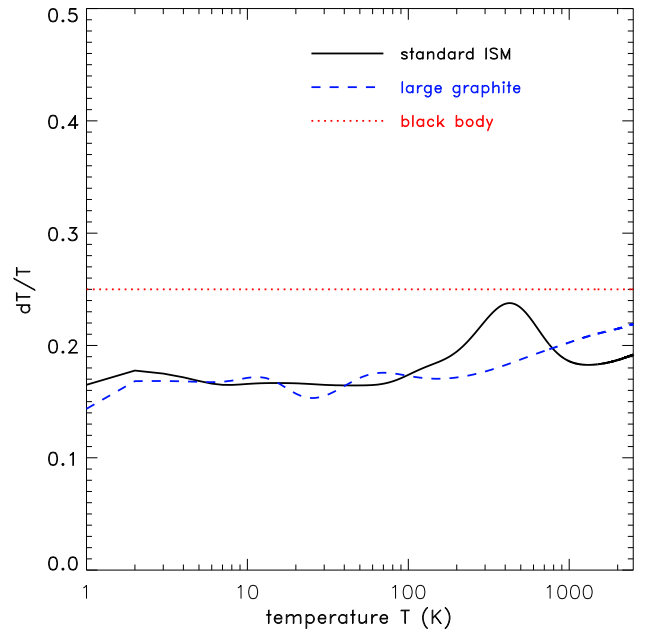}
\caption{Dependence of the relative temperature change $dT/T$ on the actual dust temperature for a variability in luminosity of $dL/L=1$. The black-solid line shows standard ISM (see text for details) while the blue-dashed line is for large graphite grains. The red-dotted line indicates the black-body limit.}\label{fig:dT}
\end{figure}

Absorption efficiencies of astronomical dust species (graphite, silicates) and compositions typically peak in the UV/optical in about the same $\nu$-range where many astronomical sources radiate (e.g. stars or AGN accretion disks). Therefore we assume for practical purposes\footnote{Without this assumption, we would have to replace $\mathrm{d}L_\mathrm{abs}/L_\mathrm{abs} = \mathrm{d}L/L \cdot Q_\mathrm{abs;\nu}/Q_\mathrm{abs;L}$ where $Q_\mathrm{abs;L} = \int L_\nu Q_\mathrm{abs;\nu} \mathrm{d}\nu \ / \ \int L_\nu \mathrm{d}\nu$ is the absorption efficiency averaged over the emission profile of the source.} $\mathrm{d}L_\mathrm{abs}/L_\mathrm{abs} \approx \mathrm{d}L/L$ (with $L = \int L_\nu \ \mathrm{d}\nu$) and obtain the variability function
\begin{equation}\label{eq:3}
\mathrm{d}T = \frac{\mathrm{d}L}{L} \cdot \left(\frac{4}{T} + \frac{Q^\prime_\mathrm{abs;P}(T)}{Q_\mathrm{abs;P}(T)}\right)^{-1} \,\, .
\end{equation}
For a given temperature $T$ of the dust (e.g. as obtained by equilibrium calculations) the change of temperature $\mathrm{d}T$ can be analytically calculated for any change in incident radiation. However the change in temperature depends on the temperature itself. On the other hand this additional $T$-dependence is minor. In Fig.~\ref{fig:dT} we show the dependence of the fractional change of temperature $dT/T$ on the dust temperature $T$ for a constant $dL/L=1$. The black-solid line shows this dependence for standard ISM dust with 53\% silicates, 47\% graphite grains \citep[optical constants from][]{Dra03}, and a grain size distribution according to \citet[][MRN]{Mat77}. For temperatures below $\sim$100\,K and above 1000\,K the fraction is nearly constant at about 0.16--0.19. Between 100\,K and 1000\,K the Planck mean opacity has a ``knee'' that is reflected in $dT/T$ as a bump up to 0.24 at 400\,K. The blue-dashed line in Fig.~\ref{fig:dT} shows large graphite grains with sizes $0.07\,\micron < a < 1\,\micron$ (MRN size distribution). Below 100\,K the graphite grain $dT/T$ wiggles around the same mean value of 0.17 as the standard ISM dust. From 100\,K to the sublimation temperature ($\sim$1500--2000\,K) it increases to about 0.2. Overall these changes are not very big and may be approximated by a constant factor. In fact, if we assume the black-body limit ($Q^\prime_\mathrm{abs;P} = 0$) we get
\begin{equation}\label{eq:4}
\frac{\mathrm{d}T}{T} = \frac{1}{4} \ \frac{\mathrm{d}L}{L} \,\, ,
\end{equation}
where the relative change in temperature is directly linked to the relative change of luminosity without any extra $T$-dependence, and the constant factor is 0.25. This is shown as a red-dashed line in Fig.~\ref{fig:dT}

It should be emphasized that this treatment is not restricted to a special class of objects (e.g. AGN, see Sect.~\ref{sec:varagn}) but can be used for dusty media in general. The presented treatment of temperature variations implicitly assumes that the reemission occurs in the optically-thin wavelength regime. Otherwise an additional ``self-heating term'' has to be included on the left side of Eq.~\ref{eq:1} and the treatment becomes iterative. However, even in case of a clumpy dust torus, where the primary emitting regions in the near- and mid-IR are the directly illuminated, optically-thin layers of otherwise optically-thick dust clouds, the temperature and emission profile is dominated by direct heating from the central source \citep{Hon10b} and the presented treatment should be approximately applicable.

\section{IR variability in AGN}\label{sec:varagn}

\subsection{Type 1 variability model}\label{sec:varmodel}

One of the applications of a variability model can be reverberation mapping measurements of AGN. Fluctuations in the incident radiation from the accretion disk, which primarily emits at UV/optical wavelengths, causes the re-emission of the dust in the torus to change over time \citep[e.g.][]{Gla04}. Since the dust is located at some distance from the accretion disk, the variability in the IR is delayed with respect to the optical emission by the light-travel time from the disk to the dust. This method has been exploited to measure the near-IR radius in nearby AGN \citep[e.g.][]{Gla04,Sug06}.

Reverberation mapping is usually done in type 1 AGN where obscuration effects are, in general, significantly lower than in type 2 AGN. In these objects the dust-/brightness distribution is projected onto a ``disk'' where the hottest dust emits at the smallest radii and the cooler dust at larger distances. \citet{Kis09a} presented a simple analytic model for the IR emission, which has been shown to be a good representation of a clumpy dust torus in type 1 AGN \citep{Hon10b}. In this concept the dust emission $S_\nu(r)$ (dominated by the optically-thin layer of an otherwise optically-thick dust cloud; see Sect.~\ref{sec:var}) at a distance $r$ from the AGN is characterized by the corresponding radiative equilibrium temperature $T$. The total emission $L(\mathrm{tor})_\nu$ of the torus is then calculated by integrating from the sublimation radius to a \citep[well selected; see][]{Hon10b} outer radius $r_\mathrm{out}$, and accounting for a radial dust distribution that is parametrized as a power law with index $\alpha$:
\begin{equation}\label{eq:5}
L(\mathrm{tor})_\nu = 4 \pi \int_{r_\mathrm{sub}}^{r_\mathrm{out}} \pi B_\nu(T(r)) \ \left(\frac{r}{r_\mathrm{sub}}\right)^{\alpha+1} \mathrm{d}r
\end{equation}
In this concept the term $(r/r_\mathrm{sub})^\alpha$ is directly related to the surface filling factor of the torus \citep[for details see the discussion in][Sect. 2.2]{Hon10b}. The dust equilibrium temperature can be obtained either by approximating a power-law \citep{Bar87} or self-consistently from Eq.~\ref{eq:1} (and using the sublimation radius and temperature as reference values) by solving the following equation:
\begin{equation}\label{eq:6}
\frac{Q_\mathrm{abs;P}(T)}{Q_\mathrm{abs;P}(T_\mathrm{sub})} \ \left(\frac{T}{T_\mathrm{sub}}\right)^4 = \left(\frac{r_\mathrm{sub}}{r}\right)^2 \,\, .
\end{equation}
For practical purposes this equation can be solved using a pre-calculated look-up table for the left term and interpolating for the respective $r$.

With these prerequisites it is possible to calculate wavelength-dependent light curves of type 1 AGN based on Eq.~\ref{eq:3}, considering the traveling time of any variability through the torus. 

\subsection{Model results} \label{sec:modres}

The model presented in Sec.~\ref{sec:varmodel} has only one parameter, $\alpha$, that represents the radial brightness distribution of the torus. We have shown that this parameter is well correlated with the actual radial distribution of the dust at least in type 1 AGN even in more complicated models \citep{Hon10b}. In consequence the radial dust distribution can be probed by observables such as the mid-IR spectral index or the comparison between mid-IR and near-IR interferometric size \citep{Hon10a,Kis11b}. Here we analyze the effect of radial brightness distribution (and, in turn, radial dust distribution) on IR light curves of AGN.

\subsubsection{Light curves}\label{sec:lc}

\begin{figure}
\includegraphics[width=0.48\textwidth]{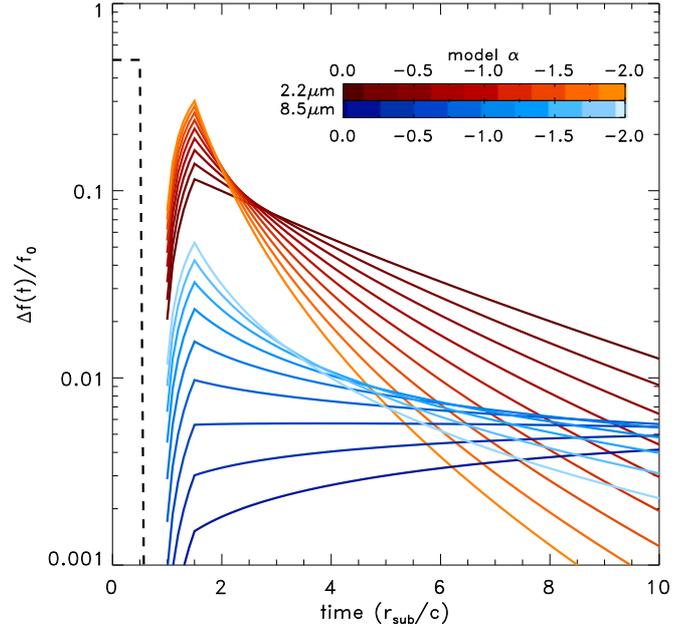}
\caption{Model light curves at $2.2\,\micron$ (red) and $8.5\,\micron$ (blue) for different values of the radial brightness distribution power law index $\alpha$. From dark to light the distribution changes from very extended to very compact. The black-dashed line illustrates the variability signal from the accretion disk.}\label{fig:lc}
\end{figure}

In Fig.~\ref{fig:lc} we show light curves $\Delta f(t)/f_0$ (where $f_0 = f(t_0)$ is the flux at time $t_0$, the starting time of the monitoring, and $\Delta f(t) = f(t)-f(t_0)$ is the flux difference between $t_0$ and $t$) at $2.2\,\micron$ and $8.5\,\micron$ for various radial brightness distributions, parametrized by the power-law index $\alpha$. The initial AGN variability signal is a step function with a width of $0.5\,r_\mathrm{sub}/c$ and an amplitude of $0.5\, \Delta L/L$, and the light curves essentially show the transfer function of this signal. The lighter colors correspond to more compact distributions. In these cases most of the light is coming from a region close to the sublimation radius, i.e. the dust is concentrated to the inner torus. The darker colors have more extended brightness distributions. Here a significant fraction of the total dust mass is located at larger radii and directly heated by the AGN. 

We plot the light curves as a function of the intrinsic time scale $t_\mathrm{sub} = r_\mathrm{sub}/c$, which corresponds to the light-travel time from the source to the sublimation radius $r_\mathrm{sub}$. In this way the variability signal from the AGN can be considered a tomographic device to map out the brightness distribution in the torus, and elapsed time and distance from the AGN become equivalent independent of the object's luminosity ($t_\mathrm{sub} = r_\mathrm{sub}/c \propto L^{1/2}$, see Sect.~\ref{sec:var}).

The near-IR ($2.2\,\micron$) light curves in Fig.~\ref{fig:lc} show that the more concentrated distributions are stronger peaked, while the distributions with $\alpha$ closer to 0 have longer tails. In the very compact objects almost all of the near-IR light comes from the peak black-body\footnote{We note that the correct description of the dust emission is a gray-body emission with source function $S_\nu = B_\nu (1-\exp(-\tau_\nu))$. For the purpose of qualitatively describing the results, however, we will use the term ``black body''.} emission of the hottest dust. When the dust distribution becomes more extended, then the Wien tails of the cooler dust emission start to contribute relatively more since the relative amount of hot dust decreases.

\begin{figure}
\includegraphics[width=0.48\textwidth]{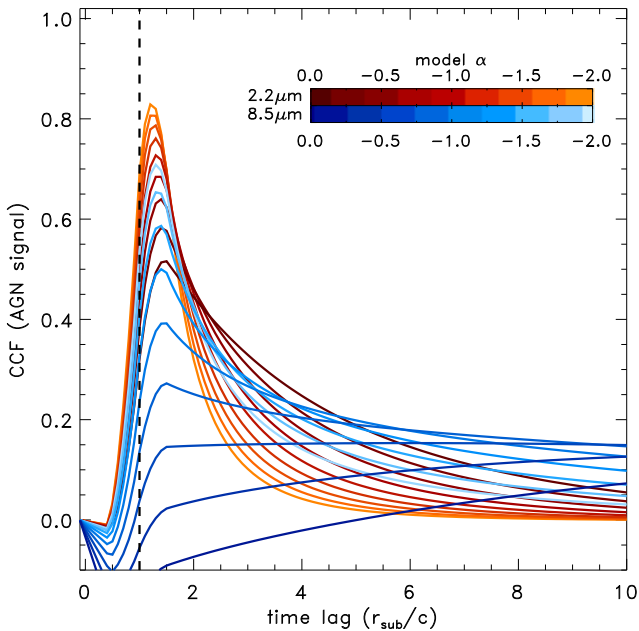}
\caption{Model cross correlation function (CCF) for light curves of AGN variability at $2.2\,\micron$ (red) and $8.5\,\micron$ (blue). From dark to light the distribution changes from very extended ($\alpha = 0.0$) to very compact ($\alpha = -2.0$). The dashed line marks a lag time of $r_\mathrm{sub}/c$.}\label{fig:ccf_agn}
\end{figure}

The behavior in the near-IR has its correspondence at longer wavelength. The mid-IR ($8.5\,\micron$) light curves with a compact brightness profile are peaked in the inner torus where the hot dust is located. Over time the brightness decays quickly, however not as fast as in the near-IR. As the brightness distribution becomes shallower (more extended), the initial peak at $r_\mathrm{sub}/c$ vanishes, the light curves get flatter, and for $\alpha>-0.5$ they keep on increasing out to $\ge10\times r_\mathrm{sub}/c$. The difference between the mid-IR and the near-IR is that in the mid-IR the emission comes predominantly from either the Rayleigh-Jeans tail of hot dust emission in the inner torus or from the peak black-body emission of cooler dust at larger distances. In the near-IR the flux originates predominantly in the black-body peak of hot dust with little contribution from the steeply dropping Wien tail of cooler dust. For steep brightness profiles, the hot-dust Rayleigh-Jeans tail dominates the mid-IR emission due to the lack of extended dust. Around $\alpha\approx -1$, the contribution of the black-body peak from extended dust becomes about equal to the hot dust Rayleigh-Jeans tail, which takes over for even shallower brightness profiles. 

\subsubsection{Lag times}

Since the contribution from different emission regions changes as a function of wavelength and brightness distribution, we can expect delays between the peak of the emission in light-curves. Classical reverberation mapping tries to quantify the lag time between the AGN variation and its correspondence in the near-IR by means of the cross correlation function (CCF)
\begin{equation}
CCF = \int f_X(t)\,f_Y(t-\Delta t) \, \mathrm{d}t
\end{equation}
where $f_X(t)$ and $f_Y(t)$ are light curves at wavebands $X$ and $Y$, and $\Delta t$ is the introduced time lag between both light curves. This method can be applied, in principle, also to the mid-IR. We note that in the framework of this paper we define lag times as the peak in the CCF.

In Fig.~\ref{fig:ccf_agn} we show the CCF of the AGN variability signal with the $2.2\,\micron$ and $8.5\,\micron$ light curves, respectively. As expected the time lag between AGN variability and near-IR emission is close to $r_\mathrm{sub}/c$ and the CCF is quite narrow. The fact that it is actually slightly larger than unity is in part a result of the used variability function (step function with width $1/2 t_\mathrm{sub}$) and the smoothened-out response by the dust. It is seen that the lag time depends slightly on $\alpha$, so that extended emission profiles show slightly longer lag times ($\sim 1.5\times r_\mathrm{sub}/c$) while the most compact brightness distributions are at $\sim 1.1 \times r_\mathrm{sub}/c$.

The mid-IR light curves show a different behavior. If the brightness distribution is compact, the peak in the CCF is still well-defined and close to $ r_\mathrm{sub}/c$. However when the distribution becomes more extended, the CCF becomes very broad. For $\alpha \le -0.5$ the peak is no longer well defined and shifts to very long time lags. This is expected from the light curves and reflects the change in contribution from hot and cooler dust to the mid-IR emission (see Sec.~\ref{sec:lc}).

\begin{figure}
\includegraphics[width=0.48\textwidth]{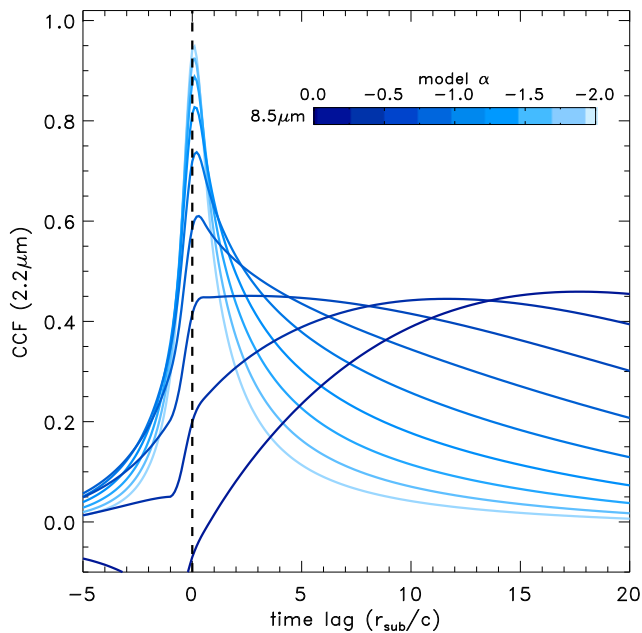}
\caption{$2.2\,\micron$--$8.5\,\micron$ model cross correlation function (CCF). From dark-blue to light-blue the distribution changes from very extended ($\alpha = 0.0$) to very compact ($\alpha = -2.0$).}\label{fig:ccf_k}
\end{figure}

The change in CCF (or light curve) with $\alpha$ may be used to distinguish between compact and extended dust distributions. Interestingly, since the near-IR CCFs and light curves are peaked around $r_\mathrm{sub}/c$ for all $\alpha$, it is not necessary to take the AGN variability signal as a reference, but instead use the near-IR as the CCF-reference signal for the mid-IR. While the actual size of $r_\mathrm{sub}$ cannot be determined in that way, the dust distribution relative to $r_\mathrm{sub}$ is still accessible. In Fig.~\ref{fig:ccf_k} we show the CCF between near-IR and mid-IR light curves. For each $\alpha$ the near-IR light curve was used as reference for the corresponding mid-IR light curve. As expected the compact brightness distributions have no time lag between near- and mid-IR, illustrating the fact that the mid-IR emission originates from the Rayleigh-Jeans tail of the hot dust emission. With $\alpha<-1$ the CCF curves become significantly wider and the lag shifts to longer time lags. For $0.0<\alpha<-0.5$ the time lags reach 5 to $20\times r_\mathrm{sub}/c$. It is therefore possible to exploit IR light curves to determine the radial brightness distributions and, in turn, the radial dust distribution as an alternative to interferometry. We note that the actual shape of the CCFs also depend on the auto correlation function of the input signal which, however, does not change the overall effects and conclusions.

\subsection{Complications when dealing with AGN light curves}\label{sec:agncomp}

One of the complications when trying to model optical and IR light curves of AGN is a possible dependence of variability on frequency. While the model uses a frequency-integrated version of $\mathrm{d}L/L$, observations are usually in one specific waveband. \citet{Meu11} report $\Delta f_\lambda \propto \lambda^{-2}$, converting to $\Delta f_\nu \propto \nu^0$ (i.e. independent of frequency), so that $\Delta f_\nu/f_\nu$ (and correspondingly $\mathrm{d}L_\nu/L_\nu$) depends on the slope of $f_\nu$. In the near-IR the slopes are assumed $f_\nu \propto \nu^{1/3}$ and probably redder towards the optical and UV \citep[e.g.][]{Zhe97,Van01,Kis08}. This means that $\Delta f_\nu/f_\nu$ in the UV is larger than in the optical, and because more energy is radiated in the UV, any light curve observed in the optical potentially underestimates the variability amplitude of the absorbed emission.\footnote{We note that this argument can also be applied to reverberation mapping of optical broad Balmer emission lines where the variability of ionizing photons is probably larger than the variability observed at optical wavelengths.} On the other hand, in the process of radiative transfer the whole accretion disk spectrum is integrated over all frequencies, convolved with the dust absorption efficiency, a process that probably smoothens out the extreme variability amplitudes. We can introduce a constant factor $w_x \equiv (\mathrm{d}L/L)/(\mathrm{d}L_x/L_x)$, the variability efficiency factor at waveband $x$, as a free parameter to compensate for these complications without requiring further assumptions on the wavelength-dependence of the variability. In this way $w_x$ will also include any effects that potentially alter the relative variability (e.g. non-variable light from the host), but are not included in this simple model.

An additional complication is a possible change in dust composition with distance from the AGN. Several studies found evidence that the inner torus is dominated by large graphite grains while the outer torus contains a mix of silicates and graphites. This leads to comparably small sublimation radii and a change of emissivity from the near-IR to mid-IR resulting in a ``bump'' at around 3\,$\micron$ in many type 1 SEDs \citep[e.g.][]{Kis07,Mor09,Kis11b}. Therefore, when comparing observed light curves in the near- and mid-IR it may become necessary to adjust the absorption efficiencies in Eqs.~\ref{eq:3} \& \ref{eq:6} accordingly.

\section{Modeling the near-IR light curve of NGC~4151}\label{sec:n4151}

A comparison between near-IR and mid-IR light curves seem to be a promising tool to constrain the brightness profiles of AGN dust tori. As of yet mid-IR light curves are very sparse. However, as described in Sect.~\ref{sec:lc}, near-IR light curves have different peak intensities and tails for compact or extended distributions with respect to the AGN variability signal. We may, thus, hope that comparing light curves in the optical and near-IR can also help decide if the brightness distribution in the torus is compact or extended, at least in the inner part of the torus.

One of the best-monitored AGN in the IR is the bright nearby Seyfert 1 galaxy NGC~4151. \citet{Kos09} presented optical and near-IR light curves with a very high temporal coverage, and the data presented below were extracted from their work. They were taken in the course of the MAGNUM project \citep{Yos03} over about 2000\,days from 2001 to 2006. We use the reduced and host- and accretion-disk-subtracted data as shown in Fig.~1 of \citet{Kos09}. As noted in \citet{Min04} the typical photometric errors for NGC~4151 data are 0.01\,mag ($\sim$1\% in flux; but see below for more details). \citet{Kos09} report that the $K$-band emission at 2.2\,$\micron$ varies similarly to the $V$-band emission with a notable delay that reflects the light-travel time from the accretion disk to the torus. Near-IR interferometry of the same object showed that the $K$-band emission is probably coming from a confined region at the dust sublimation radius \citet[i.e. consistent with a thin ring;][]{Kis09b,Kis11a}. Therefore it seems that the object is well-suited to apply our model.

The AGN variability model has essentially two parameters: the brightness distribution power law index $\alpha$ (see Sect.~\ref{sec:modres}) and the variability efficiency factor $w_V$ (see Sect.~\ref{sec:agncomp}). For the simulations we first determined the $V$-to-$K$ time lag by Monte Carlo simulations of light curves based on the observed $V$- and $K$-fluxes, following the method outlined in \citet{Sug06}. For that we first calculated the structure functions $s(t_i-t_j) = \left(\sum_{i<j}[f(t_i)-f(t_j)]^2\right)/N(i<j)$ of the $V$- and $K$-band light curves, where $f(t_i)$ is the flux at epoch $t_i$ and $N(i<j)$ is the number of $(i,j)$ pairs. From $s$ we can determine the typical change of flux for a given time interval between observations. Based on $s$ we perform Monte-Carlo simulations to interpolate the gaps in the observed light curves as follows: First we define a regular $t$-sampling $(T_1,T_2,...)$ that is finer than the observed one. Then linearly interpolated fluxes for this $t$ array are calculated based on the observed fluxes. We pick a random epoch $T_i$ from the pre-defined $t$-sampling, calculate the distance $T_i-t_\mathrm{obs}$ to the nearest observed epoch $t_\mathrm{obs}$, and determine the corresponding flux change $\Delta f_s(T_i-t_\mathrm{obs})$ from the structure function $s$. We then calculate a random flux change $\Delta f$ from the linearly interpolated flux at $T_i$ based on a Gaussian distribution with standard deviation $\Delta f_s$. The corresponding epoch and flux is added to the array of observed fluxes and the procedure repeated until fluxes are calculated for all $T_i$ in the $t$-sampling. For a proper comparison we use the same $t$-sampling for the $V$- and $K$-band light curves.

\begin{figure}
\includegraphics[width=0.48\textwidth]{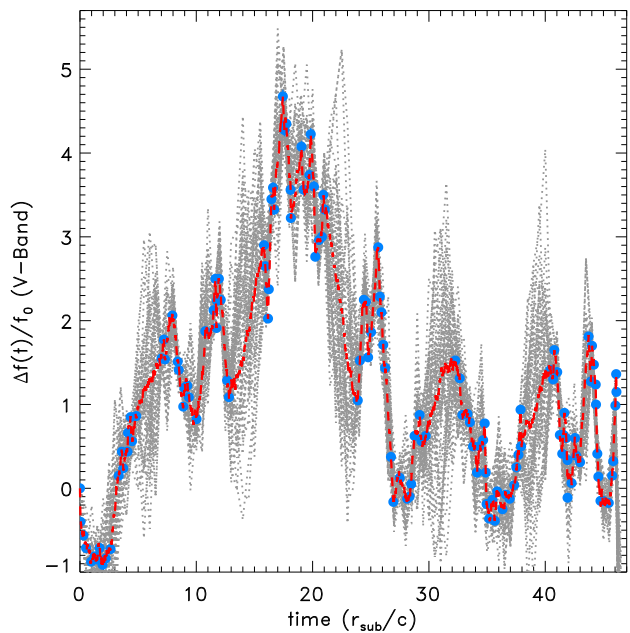}
\caption{Observed and interpolated $V$-band light curve of NGC~4151. The blue-filled circles are the observed $V$ data points. Error bars are plotted but smaller than the symbols in most cases. Based on these observations, 50 interpolated light curves have been calculated and shown as gray-dotted lines, reflecting the uncertainty of the ``true'' light curve. The red-dashed line indicates the mean of all these models which has been used as the input signal to our model simulations.}\label{fig:n4151_V}
\end{figure}

The Monte Carlo simulations are repeated 50 times to obtain an idea of the uncertainties in the light curves introduced by the finite sampling of the observations. These simulated light curves can then be used to analyze the time lag between the $V$- and $K$-band emission. We find an average time lag measured over the full light curve, excluding the strongest (and widest) peak, of $43.8\pm8.5$\,days, which corresponds to a (sublimation) radius of $r_\mathrm{sub} = 0.037\pm0.007$\,pc, in agreement with near-IR interferometry \citep{Kis09b,Pot10,Kis11a}. When using only individual features of the light curves the error is generally larger but the resulting time lag is consistent with the average one. We will discuss consequences of this seemingly constant time lag below.

For the modeling we used the average time lag as an offset between the $V$ and $K$ light curves for the model simulations. In a next step the simulated $V$-band light curves have bee used to calculate a mean $V$-band light curve. The observed and simulated $V$-band light curves, as well as the mean light curve, is shown in Fig.~\ref{fig:n4151_V}. This mean $V$-band light curve was used as the input signal for our torus variability model. 

\begin{figure}
\includegraphics[width=0.48\textwidth]{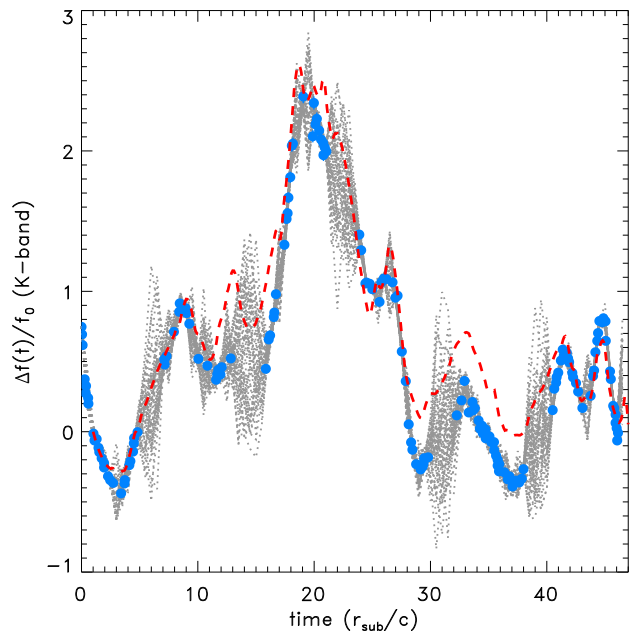}
\caption{Observed and modeled $K$-band light curve of NGC~4151. The blue-filled circles are the observed data points. Error bars are plotted but smaller than the symbols in most cases. Based on these observations, 50 interpolated light curves have been calculated and shown as gray-dotted lines, reflecting the uncertainty of the ``true'' light curve. The model light curve is overplotted as a red-dashed line.}\label{fig:n4151_mod1}
\end{figure}

\begin{figure}
\includegraphics[width=0.48\textwidth]{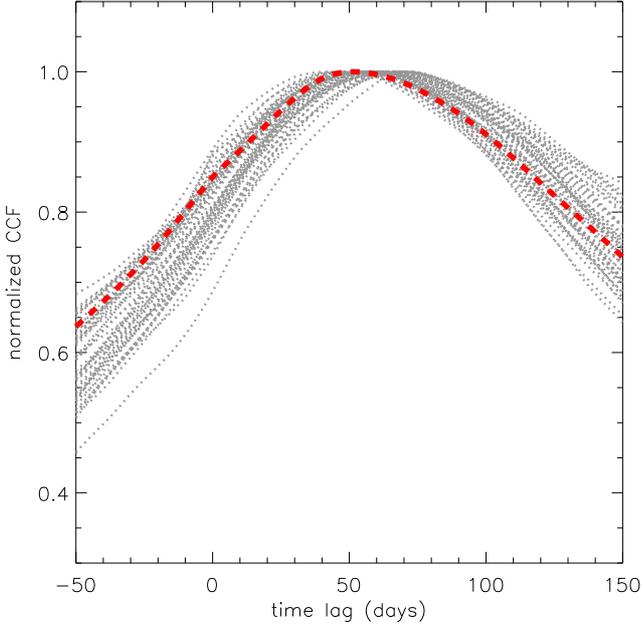}
\caption{$V$/$K$-band cross-correlation functions of NGC~4151. Based on the observed $V$- and $K$-band light curves, 50 interpolated light curves for both $V$- and $K$-band have been simulated and CCFs calculated accordingly, shown as gray-dotted lines. These reflect the observations and the uncertainty in the determination of the time lag due to the gaps and non-uniform sampling of the data. The CCF based on the light-curve model is overplotted as a red-dashed line. All CCFs have been normalized to their peak value.}\label{fig:n4151_mod1_ccf}
\end{figure}

In Figs.~\ref{fig:n4151_mod1} and \ref{fig:n4151_mod1_ccf} we compare the observed $K$-band light curve (blue-filled circles) and the CCFs based on observations (gray-dotted lines), respectively, to the best-fitting model. The blue-filled circles in Fig.~\ref{fig:n4151_mod1} are individual photometric observations extracted from \citet{Kos09}. Based on the previously described Monte Carlo interpolation scheme, we simulated 50 (nearly continuous) light curves accounting for the gaps in the observed light curve and the resulting uncertainty. They are shown as gray-dotted lines, as light curves in Fig.~\ref{fig:n4151_mod1} and as CCFs in Fig.~\ref{fig:n4151_mod1_ccf}. The range of these simulated curves reflect the range of curves if the temporal coverage would be infinite. The simulated CCFs based on the observations have been normalized so that the peak value is unity. Overplotted (red-dashed lines) is the best-fitting $K$-band model in both plots. The formal fitting was done using the light-curve only. In order to asses the quality of the fit we used two different versions of the error estimates of the observation. Although \citet{Kos09} do not provide typical errors, an initial fraction of the NGC 4151 $K$-band photometry using the same instrumentation was published in \citet{Min04} and a photometric error of about 0.01\,mag (or about 1\% in flux) was reported. When using this value for the full data set we obtain a reduced $\chi^2_\nu=4.2$. An alternative approach to estimate the error in the data comes from the structure function that we used in the Monte Carlo interpolation of the light curves. The structure function of an AGN is supposed to follow a power law from short to long intervals $t_i-t_j$ until a break at long intervals \citep[e.g.][]{Sug06}. At short time intervals the intrinsic variation should approach 0 for $t_i-t_j \rightarrow 0$. In the $K$-band data we see, however, a flattening of the structure function for $t_i-t_j \la 6$\,days, i.e. the (squared) variation of the fluxes becomes independent of the observed interval. This is more typical for uncertainty in measurements than intrinsic variability. When interpreted in this way the measurement error is 3.5\% and the best fit has $\chi^2_\nu = 1.2$.

Despite the simplicity of our model, it reproduces the overall peaks and dips of the observations and interpolations (gray region) quite well. There are, however, some notable deviations, reflected in the moderate $\chi^2_\nu$ when assuming 1\% photometric error. In particular the observed light curve drops stronger after the major outburst at 20\,$r_\mathrm{sub}/c$ leading to a lower general $\Delta f(t)/f_0$ at the peak around 33\,$r_\mathrm{sub}/c$ although the shape of this peak is well reproduced. This general flux level is only recovered at the last peak after 40\,$r_\mathrm{sub}/c$. 

\begin{figure}
\includegraphics[width=0.48\textwidth]{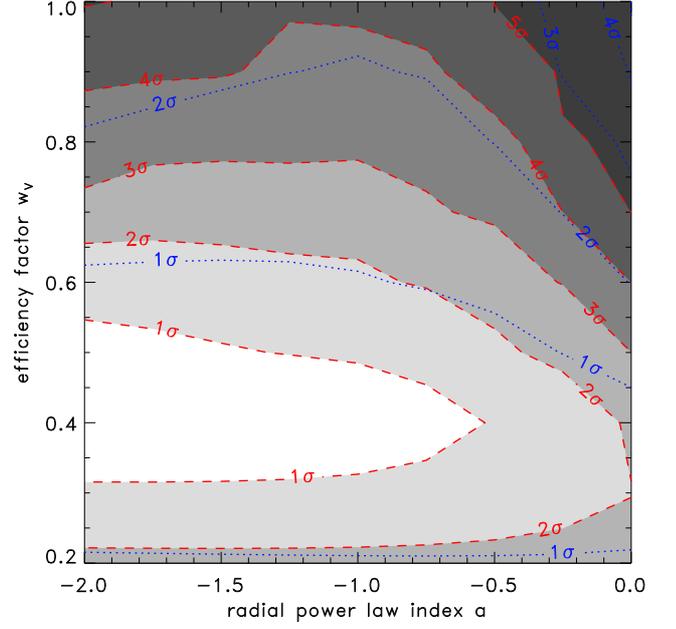}
\caption{Probability distribution of our model grid for the radial brightness distribution power law index $\alpha$ and the $V$-band efficiency factor $w_V$. The gray-shaded regions with red-dashed boundaries show the probability distribution assuming an error of the observations of 1\% from \citet{Min04} leading to $\chi^2_\nu = 4.2$. The blue-dotted lines assume an error of 3.5\% based on the analysis of the structure function with a $\chi^2_\nu=1.2$.}\label{fig:n4151_pd}
\end{figure}

The nominal best-fit model has a rather compact brightness distribution with $\alpha = -1.75$. However the probability distribution in $\alpha$ parameters is very broad ranging from about $-0.5$ to $-2.0$ (and probably beyond) with very similar $\chi^2_\nu$ values (assuming 1\% photometric error), meaning that this parameter is poorly constrained. A combination of $K$-band with longer wavelength light curves as discussed in Sect.~\ref{sec:modres} is needed to better constrain the dust distribution of the hot-dust/sublimation zone emission. The efficiency factor $w_V$ is much better constrained. In Fig.~\ref{fig:n4151_pd} we show the probability distribution for $\alpha$ and $w_V$ for both 1\% and 3.5\% photometric error. In the range of acceptable $\alpha$-values around the best fit we find $w_V \approx 0.4^{+0.2}_{-0.1}$, meaning that only about half of the power in the $V$-band variability is converted into variability in the $K$-band. Based on the discussion in Sect.~\ref{sec:agncomp} one may have expected that the $V$-band probably underestimates the amplitude of the (relative) variability of the integrated AGN emission and, therefore, leading to $w_V>1$. However a number of other factors can play a role in the value of $w_V$, e.g. uncertainties in host and accretion disk subtraction in either of the wavebands, or additional light close to the nucleus that does not participate in the variability, at least not on the covered time scales (see below for such a scenario). A more extended study using a sample of objects and/or additional UV/optical wavebands could help to better understand energy conversion from the accretion disk emission into the dust emission and its observational caveats.

Interestingly, the observed change in flux in the V-band does not seem to have shown then naively expected change of the time lag . \citet{Kos09} already note that a potential change in time lag does not scale with $(\Delta L)^{1/2}$. With a $V$-band flux change of about a factor of 20 from peak to valley\footnote{This factor seems to be a little smaller in the light curve of \citet{Sha08} covering partly the same period.}, we could have naively expected that the observed time lag/sublimation radius changed by a factor of $20^{1/2} = 4.5$ (see Eq.~\ref{eq:1}). \citet{Kos09} conclude that the time lag changes from about 60--70\,days in the first 2/3 of the light curve to about 35--50\,days in the last 1/3, although at different scaling than $(\Delta L)^{1/2}$. If such a change were significant, we should have noticed a shift in the peaks when comparing observed and modeled light curves because our models assumes a constant time lag. Over about 600\,days ($\sim13.7\,r_\mathrm{sub}/c$) the shift between observed and modeled light curve would be $\sim 6\,r_\mathrm{sub}/c$. However, such a shift is not seen in Fig.~\ref{fig:n4151_mod1}. All the peaks and valleys in modeled and observed light curve overlap within less than $1\,r_\mathrm{sub}/c$. This would have also affected the comparison of observed and modeled CCF, but both are consistent within errors (see Fig.~\ref{fig:n4151_mod1_ccf}).

In order to test even small effects of possible dust destruction and reformation, we removed the assumption of a constant time lag/sublimation radius and account for dust sublimation once it heats over the sublimation temperature (set as a free parameter) and instant or delayed reformation of the dust after cooling. This changes the sublimation radius and time lag as the variability progresses through the torus. However we find that the smallest $\chi^2_\nu$ is always found for a model with constant sublimation radius/time lag, i.e. disfavoring significant dust destruction or reformation over the observed time span. This implies that the dust can either strongly overheat before it is destroyed, or that the dust is efficiently self-shielded. The survival of a dust grain depends on the balance between gas pressure and vapor pressure. If partial gas pressure dominates, then a dust grain is stable; otherwise it evaporates. The vapor pressure of dust, $p_\mathrm{vap}\propto \exp(-1/T)$, is a strong function of the temperature $T$, while the partial pressure, $p_\mathrm{gas} \propto T$, depends only linearly on $T$. Therefore, we would expect that the lifetime of individual dust grains is short once they are heated over the sublimation temperature. As a consequence the observed behavior would favor shielding in a locally-dense environment instead of overheating of dust grains. This is consistent with the idea of a clumpy torus where the dust is confined in optically-thick clouds. A change in luminosity first acts on individual clouds which may sublimate part of their dust content but can resist longer overall at the same location than smoothly distributed dust that is not shielded locally. This scenario may also explain the low $w_V$ value: If individual clouds close sublimation radius are heated up to the sublimation temperature (and above in corresponding equilibrium temperature) and their dust gets only gradually sublimated from the surface (e.g. as a ``melting snowball''), their actual peak temperature would remain essentially constant, leading to a more or less constant $K$-band flux over some time. Hence only cooler clouds would contribute to the variability on the same time scales as the incident radiation.

\section{Summary and Conclusions}\label{sec:summary}

We present model simulations of time-variable infrared emission from dust as a consequence of variability of the incident radiation. For that we first introduce a generalized treatment for temperature variations, which can be used for all kind of dusty environments. We apply this scheme to a simplified model of a (clumpy) dusty torus around AGN and investigate how variability of the accretion disk radiation influences the torus emission in the near- and mid-IR. The main parameter of this model is the radial brightness distribution of the torus that has previously been shown to be connected to the radial distribution of the dust in the torus. We showed that any variability signal in the optical is smoothened stronger if the brightness distribution is very extended. While this effect is true for both the near- and mid-IR, longer wavelengths show much wider transfer functions than short wavelengths. The time lags between the optical and near-IR emission is mostly representing the light travel time from accretion disk to the sublimation radius independent of the brightness distribution. For mid-IR wavelengths, however, time lags can become very long, up to 10s of $r_\mathrm{sub}/c$. The effect that the brightness distribution influences the time lags seems to be much stronger than any similar effect from inclination (at least in type 1 AGN) or details of the shape of the inner torus, as recently presented by \citet{Kaw11}. This change of lag time from near-IR to mid-IR can be used to quantify the brightness distribution of the torus, either by comparing optical light curves to near-IR and mid-IR light curves, or by directly comparing near-IR to mid-IR light curves. 

Finally, the model has been applied to the optical and near-IR light curves of the nearby Seyfert 1 galaxy NGC~4151. We show that the simple model can reproduce the overall observed variability signal with some deviations in the details. For an even better match it may be necessary to use the variability scheme in the framework of more complex torus model, e.g. by incorporating it into \textit{CAT3D} \citep{Hon10b}. Nevertheless we conclude from our modeling that a single-wavelength near-IR light curve is probably not enough to constrain the brightness distribution in the torus, requiring at least one other light curve at longer wavelengths as discussed before. We found, however, that about 40\% of the energy in the variability signal in the $V$-band has been converted into $K$-band variability. This may be explained by a scenario where individual clouds close to the sublimation radius gradually sublimate their dust from the surface inward (``melting snowball''), essentially keeping their temperature constant, and, therefore, do not contribute significantly to the variability at the measured time scales. We also note that our modeling does not support a change of time lag/sublimation radius over the observed light curve epoch in spite of a significant change in $V$-band emission.

\begin{acknowledgements}
S.F.H. acknowledges support by Deutsche Forschungsgemeinschaft (DFG) in the framework of a research fellowship (``Auslandsstipendium'').
\end{acknowledgements}

\end{document}